\documentclass[preprint2]{aastex}
\newcommand{\vsini} {$v$\,sin\,$i$}
\newcommand{\kms} {km\,s$^{-1}$}
\shorttitle{$\sigma$\,Ori\,Aa, Ab, and B}
\shortauthors{Sim\'on-D\'iaz et al.}
\begin{document}
\title{A third massive star component in the $\sigma$~Orionis AB 
system\thanks{Based on observations made with the Nordic Optical Telescope, operated
on the island of La Palma jointly by Denmark, Finland, Iceland,
Norway, and Sweden, in the Spanish Observatorio del Roque de los
Muchachos of the Instituto de Astrofisica de Canarias.}}
\author{S. Sim\'on-D\'iaz} 
\affil{Instituto de Astrof\'isica de Canarias, E-38200 La Laguna, Tenerife, Spain}
\affil{Departamento de Astrof\'{\i}sica, Universidad de La Laguna, E-38205 La Laguna, Tenerife, Spain.}
\email{ssimon@iac.es}
\author{J.~A. Caballero} 
\affil{Centro de Astrobiolog\'ia (CSIC-INTA), PO Box 78, E-28691 Villanueva de la Ca\~nada, Madrid, Spain.}
\and
\author{J. Lorenzo} 
\affil{Departamento de F\'isica, Ingenier\'ia de Sistemas y Teor\'ia de la Se\~nal, Universidad de Alicante, PO Box 99, E-03080 Alicante, Spain.}
\begin{abstract}
We report on the detection of a third massive star component in the $\sigma$~Orionis AB system, 
traditionally considered as a binary system. The system has been monitored by the {\em IACOB 
spectroscopic survey of Northern Massive stars} program, obtaining 23 high-resolution
FIES@NOT spectra with a time-span of $\sim$2.5 years. The analysis of the radial velocity curves 
of the two spectroscopic components observed in the spectra has allowed us obtain the orbital 
parameters of the system, resulting in a high eccentric orbit (e\,$\sim$\,0.78) with 
an orbital period of 143.5\,$\pm$\,0.5 d. This result implies the actual presence of three stars 
in the $\sigma$~Orionis AB system when combined with previous results obtained from the study of 
the astrometric orbit (with an estimated period of $\sim$\,157\,yr).
\end{abstract}
%
\section{Introduction}\label{section.intro}

The \object{$\sigma$~Ori} star ($V \sim$ 3.80\,mag, HD~37468, HR~1931) is
visible with naked eye at about 1\,deg to the south of \object{Alnitak}
($\zeta$~Ori), the easternmost bright star of the Orion~Belt, and at a
comparable separation to the west of the \object{Horsehead Nebula}.
The actual multiple status of $\sigma$~Ori was firstly described in {\em Tabula
Nova Stellarum Duplicium}, where Mayer (1779) tabulated the stars known today as
$\sigma$~Ori~A and \object{$\sigma$~Ori~E}, separated by 42\,arcsec, and
proposed a third component in between.  
Struve et~al. (1876), from observations between 1819 and 1831, confirmed the
third component, \object{$\sigma$~Ori~D}, and reported a new ``ash''-colored one,
\object{$\sigma$~Ori~C}. 
The brightest star in the system was found to be, in its turn, a tight binary by
Burnham (1892).
The two components, historically known as \object{$\sigma$~Ori~A} and
\object{$\sigma$~Ori~B}, are thought to have very early spectral types (O9.5\,V
and B0.5\,V, respectively) and are separated by no more than 0.3\,arcsec.

The angular separations $\rho$ and $\theta$ between $\sigma$~Ori~A and B have
been frequently monitored during the last 120~years by numerous authors
with micrometers, first, and speckle imaging and adaptive optics, next. 
Some orbit determinations have been published by K\"ummritz (1958), Heintz
(1974, 1997), Hartkopf et~al. (1996), and Turner et~al (2008). 
The relatively long orbital period ($P \approx$ 157\,yr; i.e., the binary has not
completed a whole revolution since its discovery), wide projected physical
separation ($s \sim$ 100\,AU, assuming $d \approx$ 350--450\,pc) and early
spectral types have suggested that $\sigma$~Ori~A and~B may form {\em ``the most
massive visual binary known''}, with up to 40\,$M_\odot$ (Heintz 1974; Caballero
2008a). 

The small orbital semimajor axis of only 0.266$\pm$0.002\,arcsec and nearly
circular orbit (Turner et~al. 2008) makes impractical an individualized
spectroscopic study.  
Predicted radial-velocity variations in the primary have a considerable
peak-to-peak amplitude of about 7\,km\,s$^{-1}$ (Hartkopf et~al. 1996), but the 
long orbital period, magnitude difference between both components 
($\Delta H_P$ = 1.21$\pm$0.05\,mag; Perryman et~al. 1997) and scarcity of adequate 
high-resolution spectra in the last century complicate any long-term spectroscopic study.  
Some authors have reported, however, that at certain epochs the spectrum of
$\sigma$~Ori~``A--B'' changes in appearence from its single-line stage, showing
broadened and double lines separated by up to 280\,km\,s$^{-1}$ (Frost \& Adams
1904; Frost et~al. 1926; Miczaika 1950; Bolton 1974; Fullerton 1990;
Morrell \& Levato 1991; D. M. Peterson, priv. comm.). 
The time scale of these variations was only a few months.
Both observational results would not be expected from the study of the 
astrometric orbit, and could suggest that the $\sigma$~Ori~``A--B''
system is a triple system (one of the stars resolved visually being a double-lined 
spectroscopic binary, viz. Bolton 1974).
Nevertheless, the vast majority of the spectrophotometric observations of
$\sigma$~Ori~``A--B'' have not been able to resolve the spectroscopic binary
and, in words of Frost \& Adams (1904), ``evidences of complexity in the
spectrum [of $\sigma$~Ori are] scarcely sufficient to justify conclusions on the
subject''. 
More than one century later the ``subject'' is not resolved yet.

The Trapezium-like $\sigma$~Ori system being a double or a triple massive star
has major implications because of its location in the center of the homonymous
$\sigma$~Orionis cluster, which is a cornerstone for star formation studies
(Garrison 1967; Wolk 1996; B\'ejar et~al. 1999; Walter et~al. 2008; Caballero 2008b).
First, the top of the (initial) mass function and the long-term dynamical
evolution of the cluster depends on the actual mass of the brightest stars
(Sherry et~al. 2004; Caballero 2007). 
Second, any derivation of theoretical masses of intermediate- and low-mass stars
and substellar objects rely on a precise value of the cluster heliocentric
distance, which is often determined from fits of the cluster sequence in
color-magnitude diagrams (e.g., Sherry et~al. 2008; Mayne \& Naylor 2008) or via
dynamical parallax (Caballero 2008a); 
both procedures must take into account the multiplicity of $\sigma$~Ori~``A--B''.
Finally, the ``photo-erosion of pre-existing cores'' (Whitworth et al. 2007 and
references therein) by the strong high-energy radiation emitted by the
Trapezium-like system may partially explain the copiousness of cluster brown
dwarfs and planetary-mass objects (Zapatero Osorio et~al. 2000; B\'ejar et~al.
2001; Caballero et~al. 2007; Lodieu et~al. 2009; Bihain et~al. 2009), not
counting the nature of the nearby proplyd $\sigma$~Ori~IRS1~AB (van Loon \&
Oliveira 2003; Hodapp et~al. 2009) or the shape of the Horsehead Nebula (Abergel
et~al. 2003; Pound et~al. 2003; Habart et~al. 2005).  
To sum up, any investigation of the $\sigma$~Orionis cluster is incomplete if
the central $\sigma$~Ori star system is not correctly accounted~for.

\section{Observations, analysis and results}\label{section.obs+ana+res}

The $\sigma$~Ori~``A--B'' system is one of the massive binary/multiple systems
monitored by the {\em IACOB spectroscopic survey of Northern Galactic OB stars}
project (Sim\'on-D\'iaz et al. 2011). 
Using the cross-dispersed \'echelle spectrograph FIES at the Nordic Optical
Telescope, we have gathered so far 23 high-resolution (R\,=\,46,000) spectra 
with a time-span of $\sim$2.5 years (between November 2008 and April 2011; see
Table~\ref{table.rvs}). 
Our observations have been complemented with one spectrum each of $\sigma$~Ori~D
(B2\,V) and $\sigma$~Ori~E (B2\,Vpe), the other two massive star components in the
$\sigma$~Ori system.
We refer to Sim\'on-D\'iaz et~al. (2011) for a description of the spectroscopic
observations and the IACOB database. 

   \begin{table}[!t]
\footnotesize
      \caption[]{\small Radial velocity measurements of $\sigma$~Ori Aa and Ab.} 
         \label{table.rvs}
     $$ 
         \begin{tabular}{c c c c c}
            \hline
            \hline
            \noalign{\smallskip}
Date		& HJD 				& $v_{\rm rad,Aa}$		& $v_{\rm rad,Ab}$		\\
(yyyy/mm/dd)	& (--2400000)			& (km\,s$^{-1}$)		& (km\,s$^{-1}$)		\\
            \noalign{\smallskip}
            \hline
            \noalign{\smallskip}
2008/11/05 & 54776.736 &  50.1 $\pm$ 3.5 &   6.0 $\pm$ 1.9 \\
2008/11/05 & 54776.745 &  50.2 $\pm$ 3.6 &   7.0 $\pm$ 1.9 \\
2008/11/06 & 54777.697 &  50.4 $\pm$ 3.6 &   6.2 $\pm$ 2.0 \\
2008/11/06 & 54777.699 &  50.3 $\pm$ 3.6 &   6.3 $\pm$ 2.0 \\
2008/11/06 & 54777.701 &  50.4 $\pm$ 3.5 &   5.9 $\pm$ 1.9 \\
2008/11/07 & 54778.718 &  49.1 $\pm$ 3.5 &   8.0 $\pm$ 1.9 \\
2008/11/07 & 54778.720 &  49.3 $\pm$ 3.5 &   7.6 $\pm$ 2.0 \\
2008/11/08 & 54779.704 &  49.1 $\pm$ 3.5 &   6.3 $\pm$ 2.0 \\
2008/11/08 & 54779.706 &  46.9 $\pm$ 3.6 &   7.6 $\pm$ 1.9 \\
2009/11/09 & 55145.662 &  18.2 $\pm$ 3.5 &  48.0 $\pm$ 1.9 \\
2009/11/11 & 55147.650 &  14.4 $\pm$ 3.4 &  52.6 $\pm$ 1.9 \\
2010/09/07 & 55447.731 & --52.6 $\pm$ 1.2 & 128.2 $\pm$ 1.0 \\
2010/09/09 & 55449.747 & --80.3 $\pm$ 1.5 & 167.5 $\pm$ 1.2 \\
2010/10/23 & 55493.735 &  45.1 $\pm$ 3.5 &   0.6 $\pm$ 1.9 \\
2010/10/23 & 55493.737 &  45.8 $\pm$ 3.4 &   0.1 $\pm$ 1.9 \\
2011/01/11 & 55573.509 &  18.5 $\pm$ 3.6 &  40.8 $\pm$ 2.0 \\
2011/01/15 & 55577.506 &  11.3 $\pm$ 3.5 &  50.0 $\pm$ 1.9 \\
2011/01/15 & 55577.509 &  12.5 $\pm$ 3.4 &  50.0 $\pm$ 1.9 \\
2011/01/15 & 55577.512 &  11.9 $\pm$ 3.5 &  50.2 $\pm$ 2.0 \\
2011/02/11 & 55604.368 &  34.4 $\pm$ 3.5 &  28.3 $\pm$ 2.0 \\
2011/02/20 & 55613.359 &  50.3 $\pm$ 3.3 &   9.3 $\pm$ 2.0 \\
2011/03/27 & 55648.379 &  52.5 $\pm$ 3.4 &   5.1 $\pm$ 1.9 \\
2011/04/08 & 55660.353 &  47.1 $\pm$ 3.4 &   7.8 $\pm$ 2.0 \\
            \noalign{\smallskip}
            \hline
         \end{tabular}
     $$ 
   \end{table}
%

   \begin{figure}[t!]
   \centering
   \includegraphics[width=0.46\textwidth]{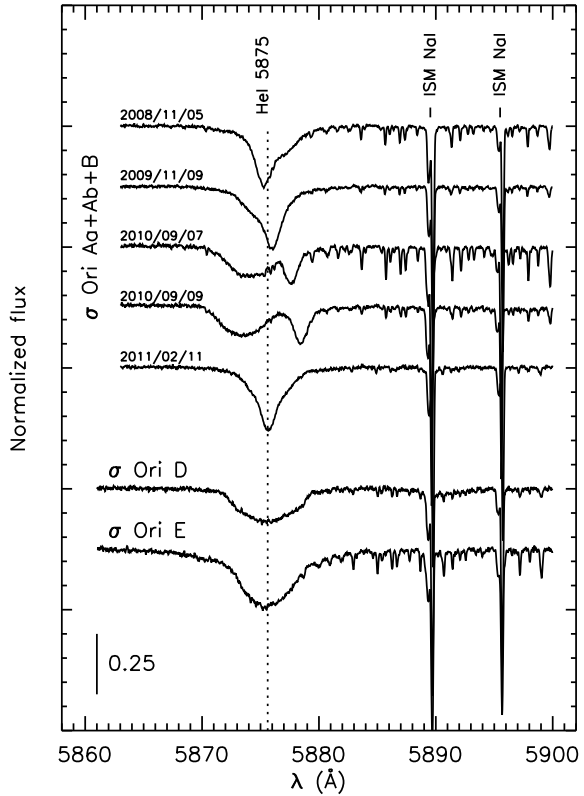}
      \caption{\footnotesize FIES spectra of $\sigma$~Ori Aa+Ab+B at five
      representative epochs (top) and of the B2\,V stars $\sigma$~Ori D and E
      (bottom) around the He~{\sc i}\,$\lambda$5875\,\AA line, corrected by
      systemic velocity and separated by 0.25 arbitrary units in normalized
      flux.  
      The sharp, deep absorption lines are interstellar Na~{\sc i}\,$\lambda
      \lambda$5890,5895\,\AA. The other shallow absorption lines are telluric.}
         \label{figure.spectra1}
   \end{figure}


Representative examples of the obtained spectra are shown in 
Fig.~\ref{figure.spectra1}, along with the spectra of $\sigma$~Ori D and E for
comparison.
At first glance, the spectra of $\sigma$~Ori~``A--B'' show two spectroscopic 
components, a narrow one and a broad one, varying on a time-scale of weeks.
The shown spectra have been corrected for heliocentric
velocity and shifted to the systemic velocity of the cluster
(+30.93\,$\pm$\,0.92\,\kms), as determined  by Sacco et al. (2008) from dozens
of FLAMES spectra of single low-mass stars in the $\sigma$~Orionis cluster. 
While the spectra of $\sigma$~Ori~D and~E show single, broad components at the
cluster systemic velocity, those of the two spectroscopic components of 
$\sigma$~Ori~``A--B'' move around that value. 

The \ion{He}{1}\,$\lambda$5875\,\AA\ line was used to determine the radial
velocity of both spectroscopic components of $\sigma$~Ori~``A--B'' for the 23
compiled spectra.  
This strong line makes easier disentagling both components and provides more
reliable radial velocity measurements than other metal lines, for which the
broad component appears too faint. 
Besides, the helium line is close to the interstellar
\ion{Na}{1}\,$\lambda\lambda$5890,5895\,\AA\ doublet, which can be used as a 
sanity check of velocity shifts and heliocentric velocity correction. 

The radial velocities were determined by means of a two-parameter cross
correlation  of the observed spectra with a grid of synthetic spectra
built with two rotationally-broadened, radial velocity-shifted ($v_{\rm rad,1}$
and $v_{\rm rad,2}$) \ion{He}{1}\,$\lambda$5875\,\AA\ line-profiles computed
with the {\sc fastwind} stellar atmosphere code (Puls et al. 2005).  
Previously, the projected rotational velocities had been determined by applying
the Fourier method (Gray 1976; see also Sim\'on-D\'iaz \& Herrero 2007 for a
recent application to OB stars) to the spectrum observed on 2010/09/09, when
both components were not blended.  
We derived \vsini\ values of 30\,$\pm$\,3 and 130\,$\pm$\,10\,\kms\ for the
narrow and broad components, respectively. 

The resulting radial velocities, together with the associated uncertainties, are
listed in Table~\ref{table.rvs}. 
The ``Aa'' and ``Ab'' components correspond to the broad- and narrow-line
spectra, respectively.


   \begin{figure}[t!]
   \centering
   \includegraphics[width=0.46\textwidth]{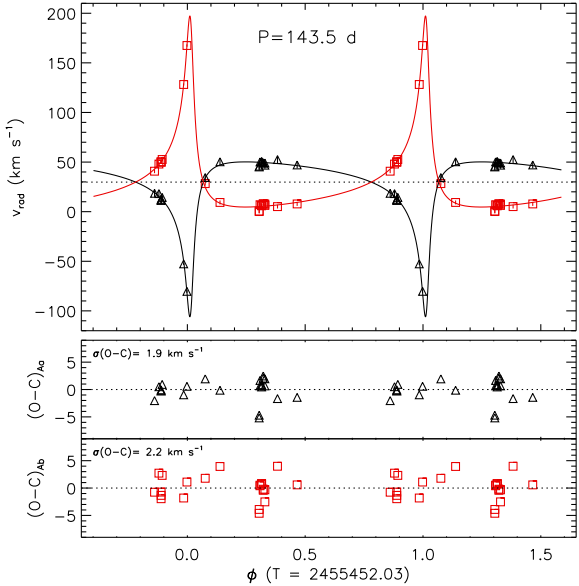}
      \caption{Upper panel: Radial velocity curves of $\sigma$~Ori Aa (triangles) and Ab
      (squares) phased to the period $P$ = 143.5\,d. 
      Errorbar sizes are smaller than those of symbol sizes. Vertical dotted line
      indicates the systemic velocity (see Table \ref{table.orbparameters}). 
      Lower panels: Velocity residuals to the adopted fit for the two components.}
         \label{figure.rv}
   \end{figure}

First, we used {\sc clean} (Roberts et~al. 1987), under an adapted version of
the original code particularly useful for unequally spaced data, for deriving a
preliminary orbital period of 147$\pm$6 days.
Next, we applied the Lehmann-Filh\'es method implemented in {\sc sbop} (Etzel
2004) for a detailed orbital analysis.  
We computed the orbital solutions for periods between $P_{\rm SB2}$ = 141.0 and
147.0\,d in steps of 0.5\,d and found that the $\chi^2$ of the radial velocity
curves of both broad and narrow components minimized at $P_{\rm SB2}$ = 143.5\,d. 
We thus assumed an error of 0.5\,d in the orbital period of the double-line
spectroscopic binary (SB2).
The final orbital solution is shown in Table~\ref{table.orbparameters} and
illustrated in Fig.~\ref{figure.rv}.
 
The derived spectroscopic orbital period, of about 
0.39\,yr, is roughly 400 times smaller than the astrometric orbital 
period of $\sim$157\,yr. Then we face a hierarchical triple system containing a close SB2 (the
components Aa and Ab) with a period of a few months and a fainter astrometric
companion (the component B) orbiting a common center of gravity with a period of
over a century. This scenario is also compatible with the good agreement found
between the center-of-mass velocity obtained for $\sigma$\,Ori Aa and Ab and the
systemic velocity determined for the cluster by Sacco et al. (2008). Following
predictions by Hartkopf et~al. (1996) both quantities should not differ by more
than $\sim$\,4 \kms.

\section{Discussion}\label{section.discussion}

   \begin{table}[t!]
\footnotesize
      \caption[]{Orbital parameters of $\sigma$~Ori Aa and Ab.} 
         \label{table.orbparameters}
     $$ 
         \begin{tabular}{l c c c}
            \hline
            \hline
            \noalign{\smallskip}
Parameter		&  \multicolumn{2}{c}{Value}					& Unit		\\
            \noalign{\smallskip}
            \hline
            \noalign{\smallskip}
$P_{\rm SB2}$		&  \multicolumn{2}{c}{143.5 $\pm$ 0.5 }				& d			\\
$T$			&  \multicolumn{2}{c}{2455452.03 $\pm$ 0.18} 	& d			\\
$e$			&  \multicolumn{2}{c}{0.7834 $\pm$ 0.012}			& 			\\
$\gamma$		&  \multicolumn{2}{c}{+29.8 $\pm$ 0.3} 			& km\,s$^{-1}$	\\
$M_{\rm Aa}/M_{\rm Ab}$	&  \multicolumn{2}{c}{1.23 $\pm$ 0.07} 		& 			\\
            \noalign{\smallskip}
            \hline
            \noalign{\smallskip}
				& Aa 			& Ab 					& 			\\
            \noalign{\smallskip}
            \cline{2-3}
            \noalign{\smallskip}
$K$			& 75 $\pm$ 4 		& 93 $\pm$ 5 				& km\,s$^{-1}$	\\
$\omega$		& 20.8 $\pm$ 0.9 &  200.8 $\pm$ 0.9			& deg		\\
$a \sin{i}$		& 132 $\pm$ 7 		& 164 $\pm$ 8 				& $R_{\odot}$	\\
$M \sin^3{i}$		& 9.3 $\pm$ 0.8 	& 7.6 $\pm$ 0.7 			& $M_{\odot}$	\\
            \noalign{\smallskip}
            \hline
         \end{tabular}
     $$ 
   \end{table}

\begin{figure*}[t!]
   \centering
   \includegraphics[height=0.99\textwidth, angle=90]{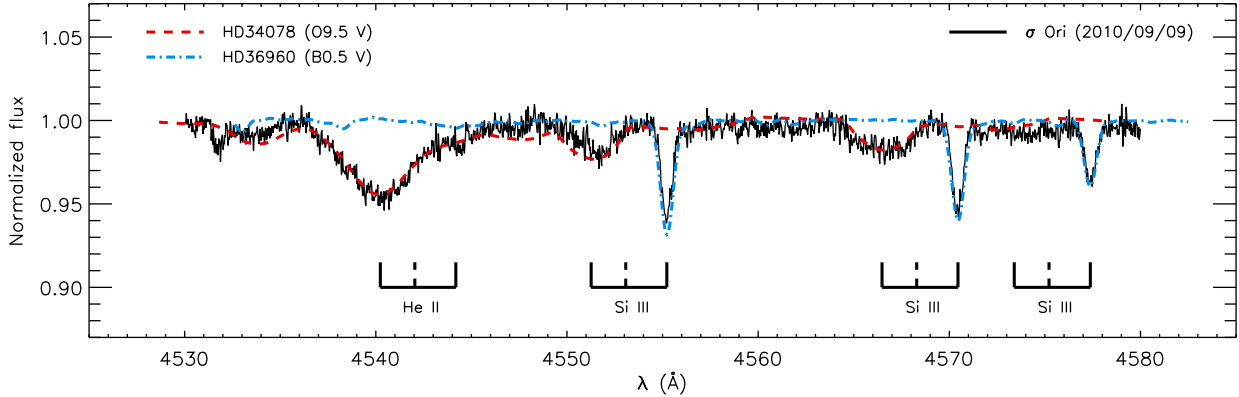}
      \caption{Spectral region showing the \ion{He}{2}\,$\lambda$4541\,\AA\ and
\ion{Si}{3}\,$\lambda\lambda$4552,4567,4574\,\AA\ lines of $\sigma$~Ori~Aa+Ab+B
at the epoch of the widest separation between components (black solid), HD~34078
(O9.5\,V; red dashed), and HD~36960 (B0.5\,V; blue dash-dotted).
The position of the lines for the components $\sigma$\,Ori Aa [left], Ab
[right], and~B [center] (measured in the case of the former ones, and expected
in the case of the latter one) are indicated in the lower part.} 
         \label{figure.spt}
   \end{figure*}

A small region of the spectrum of $\sigma$~Ori~Aa+Ab+B obtained on 2010/09/09
(corresponding to the largest separation between the broad and narrow
components) is displayed in Fig.~\ref{figure.spt}. 
The selected region contains the \ion{He}{2}\,$\lambda$4541\,\AA\ line and the
\ion{Si}{3}\,$\lambda\lambda$4552,4567,4574\,\AA\ triplet, traditionally
considered to establish the spectral type of late O- and early B-type dwarfs. 
The spectra of the standard stars HD\,34078 (\object{AE~Aur}) and
\object{HD\,36960} from the IACOB spectroscopic database (Sim\'on-D\'iaz et~al.
2011) are also shown in the figure. 
The two spectra were convolved to the projected rotational velocities
of the broad and narrow components of $\sigma$\,Ori~Aa and~Ab
(Section~\ref{section.obs+ana+res}), shifted to the radial velocities indicated
in Table~\ref{table.rvs}, and diluted by factors 0.500 and 0.275, respectively.  

Three conclusions can be made from Fig.~\ref{figure.spt}:
($i$) the two spectroscopic components are nicely fitted by the 
spectra of the O9.5\,V and B0.5\,V standards, 
($ii$) in a lower resolution spectrum (or even at a phase in which both
components are closer) the spectrum of $\sigma$\,Ori~Aa+Ab+B could be
erroneously classified as a ``single'' O9.7\,V star \citep[e.g., Sota el
al.][]{Sot11}, and ($iii$) there is no clear evidence of the B component, 
which should be located close to the systemic velocity of the 
$\sigma$\,Ori cluster, making its spectral classification difficult.
From the observed magnitude differences with respect to Aa+Ab (see e.g. 
ten~Brummelaar et al. 2000, Caballero 2008a, Ma\'iz Apell\'aniz 2010), the 
third star must be an early-B dwarf, slighly cooler and less massive.

In spite of the complexity of the system, we can estimate approximate individual
masses of the hierarchical triple system by applying the third Kepler's law  
(i.e., $(M_{\rm Aa}+M_{\rm Ab}+M_{\rm B}) P^2 = a^3$, where $a = \alpha d$ is
the physical semimajor axis in AU, $\alpha$ is the angular semimajor axis in
arcsec, $d$ is the heliocentric distance in pc, $P$ is the astrometric orbital
perior in years, and $M_i$ are the stellar masses in $M_\odot$), the ratio of
masses $M_{\rm Aa}/M_{\rm Ab}$ (Table~\ref{table.orbparameters}), and apparent
magnitudes of resolved Aa+Ab and B (see above).
Using $\alpha$ and $P$ from Hartkopf et~al. (1996) and assuming $d$ =
385$\pm$15\,pc, we estimate the individual masses at $M_{\rm Aa}$ = 19$\pm$2,
$M_{\rm Ab}$ = 15.4$\pm$1.6, $M_{\rm B}$ = 9$\pm$4\,$M_\odot$, so the
triple system contains over 40\,$M_\odot$. The system total increases to
over 60\,$M_\odot$ if we also take into account $\sigma$~Ori~C, D, and~E.


The existence of an additional powerful ultraviolet source and of a deeper
central gravity well than previously thought has a critical impact on the way
the astronomers see the $\sigma$~Orionis cluster and its surroundings. In addition,
the dynamical properties of the central, young triple system deserve further
observational and theoretical studies in the framework of the evolution 
of highly eccentric massive binaries and massive star formation theories.


\acknowledgments

{We are in debt to D.~M.~Peterson for providing us unpublished information on the 
$\sigma$~Ori star system. We thank to the anonymous referee for a very
constructive report, and to J.~Ma\'iz Apell\'aniz, R.~Barb\'a, and N.~Walborn
for a carefully reading of the manuscript before publication.
Financial support was provided by the Spanish Ministerio de Ciencia e
Innovaci\'on under projects 
AYA2008-06166-C03-01,	
AyA2008-06423-C03-03,	
and AyA2008-00695.	
This work has also been partially funded by the Spanish MICINN under the 
Consolider-Ingenio 2010 Program grant CSD2006-00070: First Science with the GTC
({\tt http://www.iac.es/consolider- \\ingenio-gtc}).}

\end{document}